\documentclass{article}
\usepackage{spconf,amsmath,graphicx}
\usepackage{amssymb, bm}
\usepackage{overpic}
\usepackage{graphicx}

\usepackage{microtype}
\usepackage{lipsum}
\usepackage{acro}
\acsetup{single}

\usepackage{comment}

\usepackage{tikz}
\usepackage{tikzscale}
\usetikzlibrary{arrows,shapes,calc,decorations.pathreplacing,intersections,quotes,angles,positioning,fit,petri,through,shadows,plotmarks,spy}
\usepackage{pgfplots}
\usepackage{subfigure}
\usepackage{import}
\usetikzlibrary{positioning}
\usetikzlibrary{3d}

\newlength{\tempdima}

\newcommand{\rowname}[1]
{\rotatebox{90}{\makebox[\tempdima][c]{\scriptsize#1}}}

\newcommand{\argmax}{\operatornamewithlimits{argmax}}
\newcommand{\argmin}{\operatornamewithlimits{argmin}}

\usepackage{enumitem}
\setlist{nosep, leftmargin=14pt}

\usepackage{mwe} 

\title{Diffusion Posterior Sampling for Synergistic Reconstruction in Spectral Computed Tomography}
\name{
	\begin{tabular}{@{}c@{}}
		Corentin Vazia$^1$, Alexandre Bousse$^3$, Béatrice Vedel$^1$, Franck Vermet$^2$, \\
        Zhihan Wang$^{3,2}$, Thore Dassow$^{3,5}$, Jean-Pierre Tasu$^{3,4}$, Dimitris Visvikis$^3$, Jacques Froment$^1$
	\end{tabular}
}

\address{$^1$ Univ Bretagne Sud, CNRS 6205, LMBA, F-56000 Vannes, France. \\
$^2$ Univ Brest, CNRS, UMR 6205, Laboratoire de Mathématiques de Bretagne Atlantique, France.\\
$^3$LaTIM  U1101, Université de Bretagne Occidentale, Brest, France.\\
$^4$Departement of Radiology, University Hospital Poitiers, France. \\
$^5$Siemens Healthcare SAS, Courbevoie, France.
}

\begin{document}
\maketitle

\begin{abstract}
Using recent advances in generative artificial intelligence (AI) brought by diffusion models, this paper introduces a new synergistic method for spectral computed tomography (CT) reconstruction. Diffusion models define a neural network to approximate the gradient of the log-density of the training data, which is then used to generate new images similar to the training ones. Following the inverse problem paradigm, we propose to adapt this generative process to synergistically reconstruct multiple images at different energy bins from multiple measurements.
The experiments suggest that using multiple energy bins simultaneously improves the reconstruction by inverse diffusion and outperforms state-of-the-art synergistic reconstruction techniques.
\end{abstract}


\section{Introduction}
For several years now, a new modality in medical imaging makes usage of the energy dependency of the attenuation with spectral computed tomography (CT). The Dual Energy CT (DUAL-CT) and Photon Counting CT (PCCT) allow us to have multiple scans at different energy levels. However, to preserve the same amount of radiation for the patient as in standard CT, each energy scan should have a lower intensity and therefore contain more noise in each measurements. The synergistic reconstruction aims to balance this fact by a complete image reconstruction of all energy images at once, using the correlation between all measurements. See for example \cite{Review_spectral_ct} for a review on this subject.

The advent of deep neural networks (DNNs) in medical imaging has opened up new perspectives by modeling the images to be reconstructed in a much more precise manner (see for example \cite{Uconnect}). In particular, generative AI models learn the structure of images from a training base  and then generate new images with similar characteristics. 
This paper considers recently introduced diffusion models \cite{Score2005,ho2020denoising} and use them as an \textit{a priori}, or regularizer, to solve ill-posed inverse problems. 
The objective is therefore to obtain a more precise modeling of images by replacing or adding to hand-crafted \textit{a priori}, such as the well-known total variation (TV) dor directional TV (DTV). We propose two diffusion model-based approaches: (i) Plain Diffusion Posterior Sampling (PDPS) which consists of reconstructing each energy bin individually (i.e, sampling from the posterior probably distribution) and (ii) Spectral DPS (SDPS) which consists in reconstruct the spectral CT images simultaneously (i.e, by sampling from the  joint  posterior probably distribution).

In Section \ref{inv_pb} and \ref{DDPM} we detail the ill-posed inverse problem of synergistic reconstruction and we present the denoising diffusion probabilistic model (DDPM) that we propose to adapt in order to solve this problem.
Implementation details and experiments are presented in Section \ref{implementation_results}. Section \ref{discussion} concludes the article and discusses future possible work.

\section{Preliminaries}\label{premilinairies}

\subsection{Synergistic Reconstruction} \label{inv_pb}

We aim to reconstruct a multichannel image $\mathbf{x}=(x^1,\dots, x^L)$, where $L$ is the number of energy bins, $x^l \in \mathbb{R}^{J}$ is the $l$-energy CT attenuation image, $J$ is the total number of pixels and $x^{l}(j)$ the corresponding linear attenuation coefficient (LAC) at pixel $j$, which is assumed not to depend on the energy within the same energy bin $l$. The available measures are $\mathbf{y} = (y^1, y^2,\dots, y^L)$ with $y^l=(y_1^l,\dots,y_I^l) \in \mathbb{R}^I$ the $l$-energy bin measure and $I$ is the total number of rays. The forward model $p(\mathbf{y}|\mathbf{x})$ we are considering is such that, for each bin $l=1,\dots,L$ and each ray $i=1,\dots,I$,
\begin{align} \label{forward_model}
 Y^l_i\mid X^l=x^l & \sim \mathcal{P}\left(\bar{y}^l_i\left(x^l\right)\right) ;\\
\bar{y}^l_i(x_l) & = h^l_i e^{-[A x^l]_i} + r_i^l, \nonumber
\end{align}
where $\mathcal{P}$ is Poisson noise applied to the expected mean $\bar{y}^l_i(x_l)$, $h^l_i$ is the total photon flux for energy bin $l$ along the ray $\mathcal{L}_{i}$,  $[A x^l]_i$ is the discretized line integral (i.e Radon transform) of the LAC $x^l$ along ray ${i}$ and $r_i^l$ is a known background term (e.g dark current). The forward model we described defines a likelihood $p(y^l|x^l)$.

On the one hand, the plain approach is to reconstruct each attenuation image $x^l$ individually from $y^l$ with the analytical method of filtered back projection (FBP) or with a dedicated iterative method, which generally seeks to solve an optimization problem.  Another classical approach would be to solve the maximum \emph{a posteriori} (MAP):
\begin{equation*}\label{eq:map}
	\hat{x}^l \in \argmax_{x^l}\, p\left(y^l|x^l\right) p\left(x^l\right) \, . 
\end{equation*}
On the other hand, synergistic reconstruction aims to solve each sub-problems simultaneously so it can use the dependencies between measurements at different energies and obtain consistent reconstructions.  The previous MAP optimization task thus becomes more general: 
\begin{equation*}\label{eq:map_multi}
	\hat{\mathbf{x}} \in \argmax_{\mathbf{x}}\, p(\mathbf{y}|\mathbf{x}) p(\mathbf{x}) \, .
\end{equation*}
In this case, we make a conditional independence assumption $p(\mathbf{y}|\mathbf{x})=\prod_l p(y^l|x^l)$ and use the same forward model \eqref{forward_model} for each $l$. The choice of the \textit{a priori} $p(\mathbf{x})$ is a difficult task as it has to encapsulate information obtained prior to the measurement. As intermediate up-to-date alternative, DTV \cite{PRISM, Synergistic_DTV} proposes to extend TV to multi channel images by enforcing structural similarities with a clean image, usually reconstructed from all energy bins. We propose here to use a generative model instead, since such models implicitly learn a prior distribution.

\subsection{Denoising Diffusion Probabilistic Model}\label{DDPM}

As shown by Song et al. \cite{song2021scorebased}, the DDPM can be seen as a discretization in $T$ steps of a stochastic differential equation (SDE) that can be reversed using a function called the score function \cite{anderson1982}. The forward diffusion model associated to this SDE progressively adds noises to the input image until obtaining (almost) pure white noise. The score function is estimated using a neural network. After training, we can generate a new sample following the target distribution by starting from white noise and by applying the reverse diffusion process. We note $p_0$ the target distribution we aim to sample from.

In detail and according to Ho et al. \cite{ho2020denoising}, given a set $(\alpha_t)_{t \in \{1,2,\dots,T\}}$ of diffusion parameters, the forward diffusion process defines a Markov process $(X_t)_{t \in \{0,1,\dots,T\}}$ starting from a clean image $x_0 \in \mathbb{R}^{J}$, assumed to be a realization of process $X_0 \sim p_0$ scaled in the range $[-1,1]$, with conditional probability distribution
\[
X_t\mid X_{t-1}=x_{t-1} \sim \mathcal{N} (\sqrt{\alpha_{t}} \ x_{t-1}, (1-\alpha_{t})I_{J}),
\]
where $I_J$ is the identity matrix in $\mathbb{R}^J$. The set of parameters $\alpha_t$ is such that, at the end of the diffusion, $X_T$ approximately follows a standard normal distribution $\mathcal{N} (0, I_{J})$. We can directly compute $x_t$ from $x_0$ with
\begin{equation}
X_t \mid X_0=x_0 \sim \mathcal{N}(\sqrt{\bar{\alpha_{t}}}x_{0}, (1-\bar{\alpha_{t}})I_{J}) \label{diff_0_to_t},
\end{equation}
where $\bar{\alpha}_t = \prod_{i=1}^t \alpha_i$. We denote by $p_t(\cdot|x_0)$ the corresponding conditional probability distribution.
The reverse diffusion process starts from $X_T \sim \mathcal{N} (0, I_{J})$ and follows a normal distribution \cite{anderson1982}:
$
X_{t-1}\mid X_t=x_t \sim \mathcal{N}\left(\mu(x_t,t), \sigma(t)^2 I_{J}\right).  \label{inverse_diffusion}
$
The generative model consists of sampling these successive probability distributions so that, starting from a realization of white noise, we obtain an image which follows the same probability distribution $p_{0}$ as the clean image $x_0$. 

We assume that the random variable $X_t$ admits a marginal density $p_t$. The  variance $\sigma(t)^2$ can be easily calculated because it only depends on $\alpha_t$, but the mean $\mu(x_t,t)$  depends on the intractable term $\nabla \mathrm{log}(p_t)(x_t)$, which is called the \textit{score function}. Using a method known as denoising score matching (DSM) \cite{Vincent2010}, the score function may be approximated by a neural network $s_{\hat{\theta}}(x_t, t)$, which makes it possible to obtain an approximation $\hat{\mu}(x_t,t)$ of the true mean $\mu(x_t, t)$:
\begin{equation} \label{mean_inverse_diffusion}
\hat{\mu}(x_t,t) \approx \mu(x_t, t) = \frac{1}{\sqrt{\alpha_t}}x_t + \frac{1-\alpha_t}{\sqrt{\alpha_t}}\nabla \mathrm{log}(p_t)(x_t).
\end{equation}
The neural network is trained on a data set $p_{\mathrm{data}}$ close to $p_0$ and such that
\[
\hat{\theta} \in \argmin_{\theta} \mathbb{E}_{t, x_0, x_t \mid x_0} \left\{ \left\Vert s_{\theta}(x_t, t) - \nabla \mathrm{log}(p_t(\cdot|x_0))(x_t) \right\Vert_2^2 \right\},
\]
where $t$ is uniformly sampled on $\{1, 2,\dots, T\}$, $x_0$ is sampled from $p_{\textrm{data}} \approx p_0$ and $x_t \mid x_0$ follows- \eqref{diff_0_to_t}.  The gradient term $\nabla \mathrm{log}(p_t(\cdot|x_0))(x_t)$ is a proxy for the true score and can be computed in closed form using \eqref{diff_0_to_t}:
\[
\nabla \mathrm{log}(p_t(\cdot|x_0))(x_t) = - \frac{x_t-\sqrt{\bar{\alpha}_t}x_0}{1-\bar{\alpha}_t}
\]
By Tweedie's formula, it is possible to show that learning the score is implicitly learning a denoising function \cite{Vincent2010}.

\begin{figure}[!ht]

	\centering
	\settoheight{\tempdima}{\includegraphics[height=0.32\linewidth]{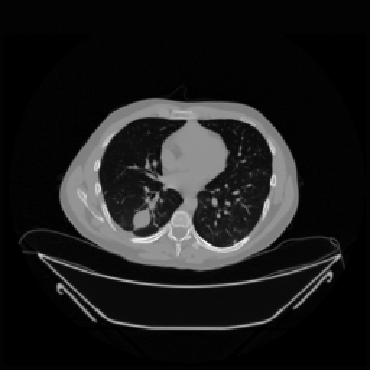}}%
	\begin{tabular}{@{}c@{}c@{\hspace{-0.2cm}}c@{\hspace{-0.2cm}}c@{\hspace{-0.0cm}}c@{\hspace{-0.0cm}}c@{}}
		& \scriptsize 40 keV & \scriptsize 80 keV & \scriptsize 120 keV  \vspace{-0.1cm} \\
		\rowname{GT} & 
       
		\begin{tikzpicture}
		\begin{scope}[spy using outlines={rectangle,yellow,magnification=2,size=7mm,connect spies}]
		\node{\includegraphics[height=\tempdima]{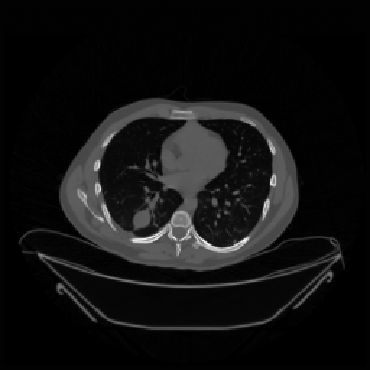}};
		\spy on (-0.1,-0.28) in node [left] at (-0.35,1.1);
		\spy on (0.3,-0.01) in node [right] at (0.35,1.1);     
		\end{scope}
		\end{tikzpicture} 
  
		&
		\begin{tikzpicture}
		\begin{scope}[spy using outlines={rectangle,yellow,magnification=2,size=7mm,connect spies}]
		\node {\includegraphics[height=\tempdima]{Res/GT_80kev.png}};
		\spy on (-0.1,-0.28) in node [left] at (-0.35,1.1);
		\spy on (0.3,-0.01) in node [right] at (0.35,1.1);     
		\end{scope}
		\end{tikzpicture} 
		&
		\begin{tikzpicture}
		\begin{scope}[spy using outlines={rectangle,yellow,magnification=2,size=7mm,connect spies}]
		\node {\includegraphics[height=\tempdima]{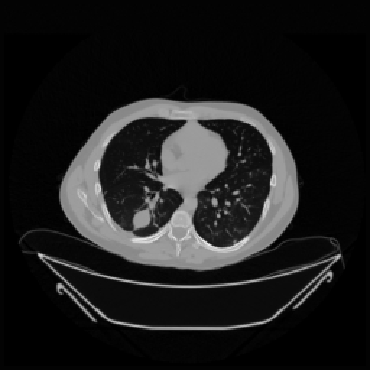}};
		\spy on (-0.1,-0.28) in node [left] at (-0.35,1.1);
		\spy on (0.3,-0.01) in node [right] at (0.35,1.1);     
		\end{scope}
		\end{tikzpicture} 
		\vspace{-0.16cm}\\
		
		\rowname{FBP} & 
		\begin{tikzpicture}
		\begin{scope}[spy using outlines={rectangle,yellow,magnification=2,size=7mm}]
		\node {\includegraphics[height=\tempdima]{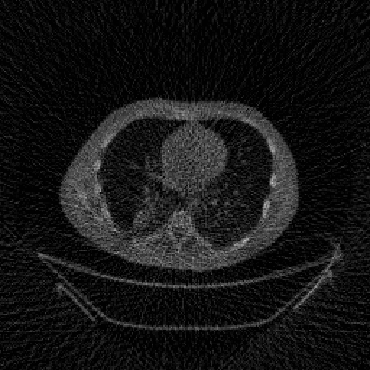}};
        \node[red, align=center] at (0,-0.70){\footnotesize PSNR = 20.71};
        \node[red, align=center] at (0,-0.95) {\footnotesize SSIM = 0.356};
		\spy on (-0.1,-0.28) in node [left] at (-0.35,1.1);
		\spy on (0.3,-0.01) in node [right] at (0.35,1.1);   
		\end{scope}
		\end{tikzpicture}
		&
		\begin{tikzpicture}
		\begin{scope}[spy using outlines={rectangle,yellow,magnification=2,size=7mm}]
		\node {\includegraphics[height=\tempdima]{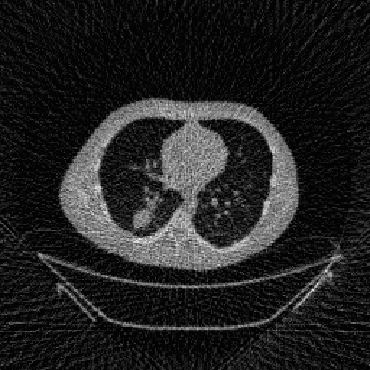}};
        \node[red, align=center] at (0,-0.70){\footnotesize PSNR = 21.30};
        \node[red, align=center] at (0,-0.95) {\footnotesize SSIM = 0.362};
		\spy on (-0.1,-0.28) in node [left] at (-0.35,1.1);
		\spy on (0.3,-0.01) in node [right] at (0.35,1.1);     
		\end{scope}
		\end{tikzpicture} 
		&
		\begin{tikzpicture}
		\begin{scope}[spy using outlines={rectangle,yellow,magnification=2,size=7mm}]
		\node {\includegraphics[height=\tempdima]{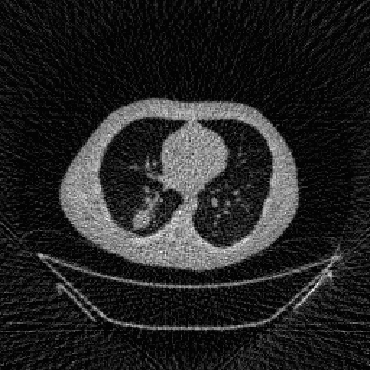}};
        \node[red, align=center] at (0,-0.70){\footnotesize PSNR = 21.24};
        \node[red, align=center] at (0,-0.95) {\footnotesize SSIM = 0.362};
		\spy on (-0.1,-0.28) in node [left] at (-0.35,1.1);
		\spy on (0.3,-0.01) in node [right] at (0.35,1.1);     
		\end{scope}
		\end{tikzpicture} 
		\vspace{-0.16cm}\\		
		\rowname{PDPS} & 
		\begin{tikzpicture}
		\begin{scope}[spy using outlines={rectangle,yellow,magnification=2,size=7mm}]
		\node {\includegraphics[height=\tempdima]{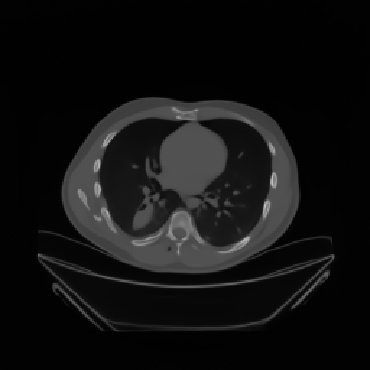}};
        \node[red, align=center] at (0,-0.70){\footnotesize PSNR = 23.41};
        \node[red, align=center] at (0,-0.95) {\footnotesize SSIM = 0.737};
		\spy on (-0.1,-0.28) in node [left] at (-0.35,1.1);
		\spy on (0.3,-0.01) in node [right] at (0.35,1.1);  
		\end{scope}
		\end{tikzpicture}
		&
		\begin{tikzpicture}
		\begin{scope}[spy using outlines={rectangle,yellow,magnification=2,size=7mm}]
		\node {\includegraphics[height=\tempdima]{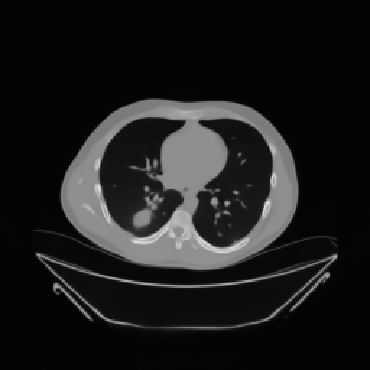}};
        \node[red, align=center] at (0,-0.70){\footnotesize PSNR = 26.03};
        \node[red, align=center] at (0,-0.95) {\footnotesize SSIM = 0.802};
		\spy on (-0.1,-0.28) in node [left] at (-0.35,1.1);
		\spy on (0.3,-0.01) in node [right] at (0.35,1.1);     
		\end{scope}
		\end{tikzpicture} 
		&
		\begin{tikzpicture}
		\begin{scope}[spy using outlines={rectangle,yellow,magnification=2,size=7mm}]
		\node {\includegraphics[height=\tempdima]{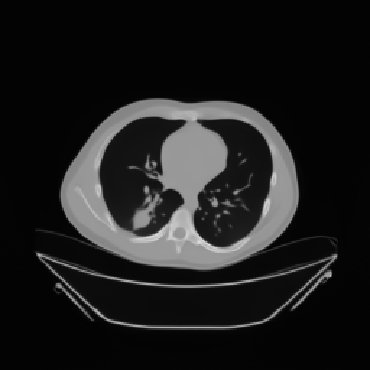}};
        \node[red, align=center] at (0,-0.70){\footnotesize PSNR = 26.59};
        \node[red, align=center] at (0,-0.95) {\footnotesize SSIM = 0.819};
		\spy on (-0.1,-0.28) in node [left] at (-0.35,1.1);
		\spy on (0.3,-0.01) in node [right] at (0.35,1.1);     
		\end{scope}
		\end{tikzpicture} 
		\vspace{-0.16cm}\\		
		\rowname{SDPS} & 
		\begin{tikzpicture}
		\begin{scope}[spy using outlines={rectangle,yellow,magnification=2,size=7mm}]
		\node {\includegraphics[height=\tempdima]{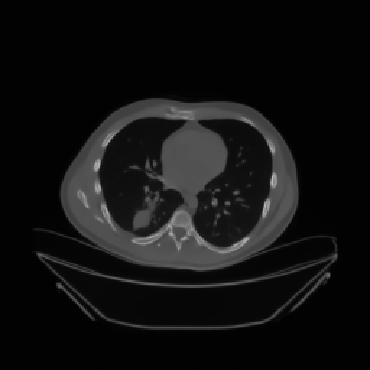}};
        \node[red, align=center] at (0,-0.70){\footnotesize PSNR = 26.93};
        \node[red, align=center] at (0,-0.95) {\footnotesize SSIM = 0.851};
		\spy on (-0.1,-0.28) in node [left] at (-0.35,1.1);
		\spy on (0.3,-0.01) in node [right] at (0.35,1.1);  
		\end{scope}
		\end{tikzpicture}
		&
		\begin{tikzpicture}
		\begin{scope}[spy using outlines={rectangle,yellow,magnification=2,size=7mm}]
		\node {\includegraphics[height=\tempdima]{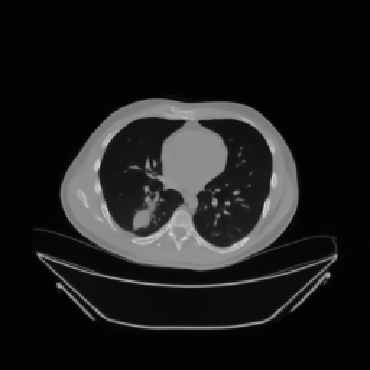}};
        \node[red, align=center] at (0,-0.70){\footnotesize PSNR = 28.19};
        \node[red, align=center] at (0,-0.95) {\footnotesize SSIM = 0.850};
		\spy on (-0.1,-0.28) in node [left] at (-0.35,1.1);
		\spy on (0.3,-0.01) in node [right] at (0.35,1.1);     
		\end{scope}
		\end{tikzpicture} 
		&
		\begin{tikzpicture}
		\begin{scope}[spy using outlines={rectangle,yellow,magnification=2,size=7mm}]
		\node {\includegraphics[height=\tempdima]{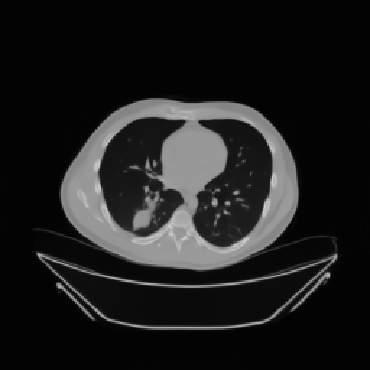}};
        \node[red, align=center] at (0,-0.70){\footnotesize PSNR = 28.05};
        \node[red, align=center] at (0,-0.95) {\footnotesize SSIM = 0.847};
		\spy on (-0.1,-0.28) in node [left] at (-0.35,1.1);
		\spy on (0.3,-0.01) in node [right] at (0.35,1.1);     
		\end{scope}
		\end{tikzpicture} 
		\vspace{-0.16cm}\\

		\rowname{DTV} & 
		\begin{tikzpicture}
		\begin{scope}[spy using outlines={rectangle,yellow,magnification=2,size=7mm}]
		\node {\includegraphics[height=\tempdima]{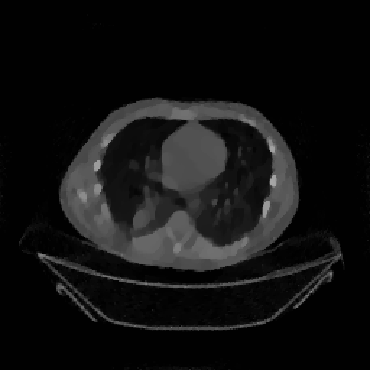}};
        \node[red, align=center] at (0,-0.70){\footnotesize PSNR = 24.55};
        \node[red, align=center] at (0,-0.95) {\footnotesize SSIM = 0.780};
		\spy on (-0.1,-0.28) in node [left] at (-0.35,1.1);
		\spy on (0.3,-0.01) in node [right] at (0.35,1.1);   
		\end{scope}
		\end{tikzpicture}
		&
		\begin{tikzpicture}
		\begin{scope}[spy using outlines={rectangle,yellow,magnification=2,size=7mm}]
		\node {\includegraphics[height=\tempdima]{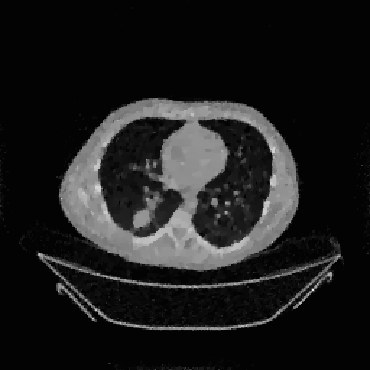}};
        \node[red, align=center] at (0,-0.70){\footnotesize PSNR = 27.56};
        \node[red, align=center] at (0,-0.95) {\footnotesize SSIM = 0.782};
		\spy on (-0.1,-0.28) in node [left] at (-0.35,1.1);
		\spy on (0.3,-0.01) in node [right] at (0.35,1.1);     
		\end{scope}
		\end{tikzpicture} 
		&
		\begin{tikzpicture}
		\begin{scope}[spy using outlines={rectangle,yellow,magnification=2,size=7mm}]
		\node {\includegraphics[height=\tempdima]{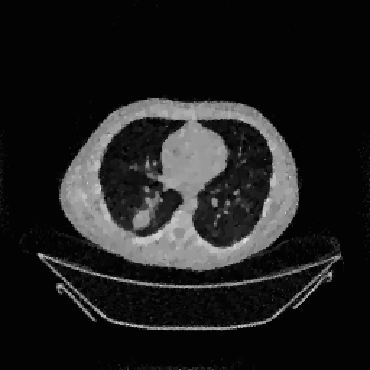}};
        \node[red, align=center] at (0,-0.70){\footnotesize PSNR = 27.14};
        \node[red, align=center] at (0,-0.95) {\footnotesize SSIM = 0.750};
		\spy on (-0.1,-0.28) in node [left] at (-0.35,1.1);
		\spy on (0.3,-0.01) in node [right] at (0.35,1.1);     
		\end{scope}
		\end{tikzpicture} 
        \vspace{-1.9cm}\\
        
		\rowname{} & 
		\begin{tikzpicture}
		\node {\includegraphics[width=\tempdima]{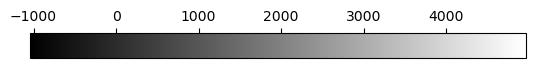}};
		\end{tikzpicture}
		&
		\begin{tikzpicture}
		\node {\includegraphics[width=\tempdima]{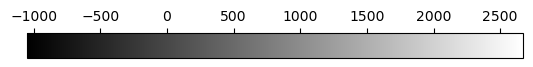}};
		\end{tikzpicture} 
		&
		\begin{tikzpicture}
		\node {\includegraphics[width=\tempdima]{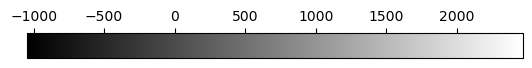}};
		\end{tikzpicture} 
		
	\end{tabular}	
	\caption{Reconstruction results of a $3$ energy bins scan with fan beam geometry and $120$ angles. Each column represents an energy of $40$, $80$ and $120$keV respectively. First line is the ground truth and other lines are the different reconstruction methods. Pixel values are in Hounsfield Units (HU). The figure also provides peak signal-to-noise ratio (PSNR) and structural similarity (SSIM) for each channels and methods. Metrics are computed on a rectangle containing the patient. We set $h_i^l = 2500 \: \forall k,l$.}\label{fig:res_images}
\end{figure}

\begin{figure*}
    \centering
    \includegraphics[width=0.9\textwidth]{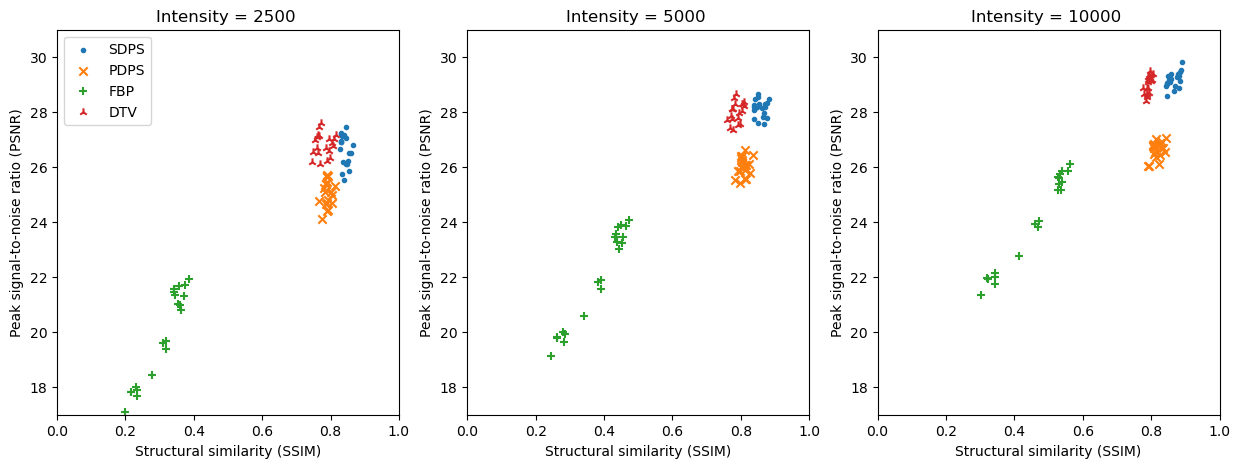}
    \caption{Reconstruction results for 30 different    slices. For each plot, we represent the peak signal-to-noise ratio on the y-axis and the structural similarity on the x-axis. The third plot is when the source intensity is high enough to obtain good analytical reconstruction with the filtered back projection. The other two plots use a lower intensity of the source that leads to more noise in the measures. We can observe that both DPS methods are robust to this increase of noise.}
    \label{fig:res_multi_tests}
\end{figure*}

\subsection{Diffusion Posterior Sampling}\label{method}

Diffusion models have already been used to solve inverse problems, see for instance \cite{song2022solving, chung2022improving, liu2022dolce}. In this context, the generative model benefits from conditioning the reverse diffusion to measurements $y$ and we then obtain a \textit{conditional score function} $\nabla \mathrm{log}(p_t(\cdot|y))(x_t) $, which is then used to sample an image $x$ from the posterior probability $p(y|x) p(x)$. 

To avoid supervised learning, the conditional score can be approximated. We use the diffusion posterior sampling (DPS) \cite{chung2023diffusion} method which allows unsupervised learning with respect to observations. This method first decomposes the conditional score function using Bayes rule:
\[
\nabla \mathrm{log}(p_t(\cdot | y))= \nabla \mathrm{log}(p_t) +\nabla  \mathrm{log}(p_t(y|\cdot)).
\]
The first term $\nabla \mathrm{log}(p_t) $ is the unconditional score function and is learned in an unsupervised manner with DSM as previously stated. The second term $\nabla  \mathrm{log}(p_t(y|\cdot))$ is approximated by:
\begin{equation} \label{approx_log_likelihood}
\nabla  \mathrm{log}(p_t(y|\cdot)) \approx \nabla  \mathrm{log}(p(y|\hat{x}_0(\cdot,t))),
\end{equation}
where $p(y|x_0)$ is the forward model \eqref{forward_model} described  in Section~\ref{inv_pb} and $\hat{x}_0(\cdot,t): \mathbb{R}^J \rightarrow \mathbb{R}^J $ is an approximation of the clean image $x_0$ from the current noisy image $x_t$ using Tweedie's formula and the unconditional score:
\begin{equation} \label{eq:tweedie}
\hat{x}_0(x_t,t) = \frac{x_t + (1-\bar{\alpha}_t) \nabla \mathrm{log}(p_t)(x_t)}{\sqrt{\bar{\alpha}_t}}.  
\end{equation}
The term $\mathrm{log}\left(p(y|\hat{x}_0(x_t,t))\right)$ in \eqref{approx_log_likelihood} is an approximation of the log-likelihood $\mathrm{log}\left(p_{t}(y|x_t)\right)$. 
We refer to \cite{chung2023diffusion} for more details. Note that, the latter being computed by the neural network, the application of the gradient along $x_t$ requires automatic differentiation.
Finally, we simply plug the approximated conditional score function into \eqref{mean_inverse_diffusion} and obtain the sampling procedure to solve the inverse problem:
\begin{multline} \label{eq:dps}
        x_{t-1} = \underbrace{\frac{1}{\sqrt{\alpha_t}}x_t + \frac{1-\alpha_t}{\sqrt{\alpha_t}} s_{\hat{\theta}}(x_t,t) + \sigma(t)z_t }_{\textrm{Unconditional inverse diffusion}} \\
    + \underbrace{\frac{1-\alpha_t}{\sqrt{\alpha_t}} \nabla \mathrm{log}\Bigl(p(y|\hat{x}_0(\cdot,t))\Bigl)(x_t)}_{\textrm{Pseudo conditional guidance}}.
\end{multline}
For all $y=y^l$ and $x=x^l$ in the previous equation, we can train $L$ neural networks and use those for plain reconstruction of each energy bin. We name this method Plain Diffusion Posterior Sampling (PDPS).

Similarly, synergistic reconstruction can be achieved by reverse sampling of a collection of spectral images $(\mathbf{x}_1,\dots,\mathbf{x}_T)$,  $\mathbf{x}_t = \left( x_t^l  \right)_{l=1,\dots,L}$, from the measurement $\mathbf{y}=(y^1,y^2,\dots,y^L)$, to sample images according to the joint posterior $p(\mathbf{y}|\mathbf{x})p(\mathbf{x})$
as follows:
\begin{multline*} \label{eq:sdps}
	\mathbf{x}_{t-1} = \underbrace{\frac{1}{\sqrt{\alpha_t}}\mathbf{x}_t + \frac{1-\alpha_t}{\sqrt{\alpha_t}} \mathbf{s}_{\hat{\bm{\theta}}}(\mathbf{x}_t,t) + \sigma(t)\mathbf{z} }_{\textrm{Unconditional synergistic inverse diffusion}} \\
	+ \underbrace{\frac{1-\alpha_t}{\sqrt{\alpha_t}} \nabla \log\left(p\left(\mathbf{y}|\hat{\mathbf{x}}_0(\cdot,t)\right)\right)\left(\mathbf{x}_t\right)}_{\textrm{Pseudo conditional guidance}} \, .
\end{multline*}
where $\mathbf{z} \sim \mathcal{N}(0,I_{J\times L})$ and $\hat{\mathbf{x}}_0 $ is the approximated clean multi-energy image from \eqref{eq:tweedie}. The DNN $\mathbf{s}_{\hat{\bm{\theta}}}(\mathbf{x}_t,t)$ is an approximation of the multi-energy conditional score  $\nabla  \log(p_t(\mathbf{x}_t))$ trained as 
\begin{equation*}
	\hat{\bm{\theta}} \in \argmin_{\bm{\theta}} \, \mathbb{E}_{t, \mathbf{x}_0, \mathbf{x}_t \mid \mathbf{x}_0}\!  \left\Vert \mathbf{s}_{\bm{\theta}}(\mathbf{x}_t, t) - \nabla\! \log(p_t(\cdot|\mathbf{x}_0))(\mathbf{x}_t) \right\Vert_2^2,
\end{equation*}  
where the expectation is computed by drawing the entire multi-energy images $\mathbf{x}_0 = (x_0^1,\dots,x_0^L)$ from the training dataset.
We name this method Spectral DPS (SDPS).
Please note that all neural networks are completely agnostic from the synergistic reconstruction task and can be used for any inverse problems regarding spectral CT.

\section{Implementation and Results}\label{implementation_results}

Numerical experiments are made on data coming from Poitiers University Hospital, France. All images are linearly rescaled to $[-1, 1]$ for the diffusion process. However, in order to retrieve LAC in pixel$^{-1}$, we compute the mean of a scout reconstruction (e.g FBP) and use it to rescale the images back into LAC. In the following, we use $L=3$ energy bins. 
For both plain and spectral DPS, the networks architecture is a simple U-Net \cite{u_net}, with sinusoidal positional time $t$ embedding \cite{vaswani2017attention}. All networks share the same numbers of epochs, batch size and learning rate schedule. They are trained on $3550$ multi-energies images ($40$~keV, $80$~keV and $120$~keV) sliced from $10$ patients and rescaled to a size of $256\times{}256$, with LAC converted to pixel$^{-1}$. Reconstruction tests are made on slices from another patient which is not part of the training set.
In order to use automatic differentiation form Pytorch library, we use the \textit{torch-radon} module \cite{torch_radon} to perform projections and back-projections. We follow the DPS method \cite{chung2023diffusion} and normalize the gradient of the pseudo conditional guidance in \eqref{eq:dps} and use a fixed gradient descent step of $0.7$. Finally, we only operate the reverse-diffusion on the last $T' = 100$ steps instead of the all $T=1000$ steps by starting the reverse diffusion process on a diffused scout reconstruction instead of white noise. This implementation is closely related to the contraction theory of the stochastic differential equation \cite{chung2022come} that, once discretized, result to the DDPM. Even if the training procedure of the neural network can be time-consuming, the inference is fast and takes around $10$ seconds for the reconstruction of one slice on a NVIDIA GeForce RTX 2080 Ti GPU.


 We compare both SPDS and PDPS methods to the baseline FPB reconstruction and the state of the art synergistic reconstruction with DTV \cite{Synergistic_DTV}, which uses a guidance image to reconstruct each energy image. The directional total variation method is performed using a forward-backward splitting algorithm as in \cite{Synergistic_DTV} with $500$ iterations for each reconstruction. 
Fig.~\ref{fig:res_images} shows the results of reconstruction from a $120$ angles fan beam projection measurement with low intensity. We used a test data set and projected each image with Torch-Radon, and then applied the forward model described in Section \ref{inv_pb}.  Although the PDPS method produces good results, we observe coherence loss between the different energy images. In contrast, the SDPS method maintains coherence between images, and details found in one channel can also be seen in another. 
Fig.~\ref{fig:res_multi_tests} presents results from multiple reconstruction of $20$ slices from all methods presented here and at different level of source intensity. We notice that while SDPS and DTV offer similar PSNR results, the former method gives better SSIM and coherence between energy bins as well as a finer detail reconstruction than DTV, see Figure \ref{fig:res_images}.


\section{Conclusion and Future Work}\label{discussion}
We  proposed a diffusion method named SDPS that solves the inverse problem of synergistic reconstruction. We showed empirically on real data that SDPS is efficient and produces coherent images, even when the noise in the measurements is high. Further work might be to use a diffusion model that is more suited to the Poisson noise of the forward model instead of the Gaussian diffusion and/or use this efficient reconstruction method to perform direct multi-material decomposition. 

\section{Acknowledgments}
\label{sec:acknowledgments}
This work was conducted within the France 2030 framework programmes, Centre Henri Lebesgue ANR-11-LABX-0020-01 and French National Research Agency (ANR) under grant No ANR-20-CE45-0020, and with the support of \textit{Région Bretagne}. This research study was conducted retrospectively using human subject data, ethical approval was not required. All authors declare that they have no known conflicts of interest in terms of competing financial interests or personal relationships that could have an influence or are relevant to the work reported in this paper.

\bibliographystyle{IEEEbib}
\bibliography{biblio}

\end{document}